\documentstyle[12pt,graphicx,amssymb]{article}

\begin{document}

\title{Low-density series expansions for directed percolation II:\\
The square lattice with a wall}

\author{Iwan Jensen\thanks{e-mail: I.Jensen@ms.unimelb.edu.au} \\
Department of Mathematics and Statistics, \\
University of Melbourne, Parkville, Victoria 3052, Australia}

\maketitle

\begin{abstract}
A new algorithm for the derivation of low-density expansions
has been used to greatly extend the series for moments of the 
pair-connectedness on the directed square lattice near an 
impenetrable wall. Analysis of the series yields very accurate
estimates for the critical point and exponents. In particular, 
the estimate for the exponent characterizing the average cluster 
length near the wall, $\tau_1=1.00014(2)$, appears to exclude the
conjecture $\tau_1=1$. The critical point and the exponents 
$\nu_{\parallel}$ and $\nu_{\perp}$ have the same values as 
for the bulk problem.
\end{abstract}


\section{Introduction \label{sec:intro}}

Surface critical behaviour in equilibrium systems has received a lot 
of attention in recent years \cite{Binder,Diehl}. Close to a surface, 
thermodynamic quantities have critical exponents which may depend on 
the specific boundary conditions and differ from the exponents 
characterizing the behaviour in the bulk. While relatively little
attention has been paid to surface critical behaviour in non-equilibrium 
systems, it is clear that, as in the bulk case, similar scaling ideas and
principles should apply \cite{Lauritsen98}. 

Directed bond percolation (DP) \cite{DPreview,PercTh} can be thought of 
as a purely geometric model in which bonds placed on the edges of a 
lattice are either present with probability $p$ or absent with probability 
$q=1-p$. Connections are only allowed along a preferred direction 
given by an orientation of the edges. The behaviour of the model is 
controlled by the branching probability (or density of bonds) $p$. When 
$p$ is smaller than a critical value $p_c$ all clusters of connected bonds
remain finite. Above $p_c$ there is a non-zero probability of finding an 
infinite cluster, and the average cluster size $S(p)$ diverges as 
$p \rightarrow p_c$. In most interpretations the  preferred direction is 
time, and one realization is as a model of an epidemic without immunization. 
More precisely, the directed square lattice has sites which are the points 
in the $t - x$ plane with integer co-ordinates such that $t \ge 0$ and 
$t + x$ is even. There are two edges leading from the site $(t,x)$ which 
terminate at the sites $(t+1, x \pm 1)$. A wall will be said to be present 
if the bonds leading to sites with $x < 0$ are always absent. In a recent 
paper \cite{EGJT} this model was studied using series expansions. The 
analysis of the series, which were calculated to order 67, raised the very 
interesting possibility that the mean cluster length with the initial seed 
close to the wall has a critical exponent $\tau_1 =1$. Here, the subscript 
1 is used to  indicate that the value of the exponent is in the presence of 
a wall. The value of $\tau_1$ has also been obtained from Monte Carlo 
simulations \cite{Lauritsen97} and the estimate  $\tau_1 =1.000(5)$ is 
consistent with the conjecture $\tau_1 =1$. If true this would be quite 
remarkable since no other (non-trivial) exponents for the DP problem appear 
to be given by integers or even (simple) rational fractions 
\cite{JG,Jensen96}.

In an effort to further investigate this problem I have used a recently 
devised and very efficient algorithm \cite{Jensen99} to greatly extend 
the low-density series for moments of the pair-connectedness. The series 
have been derived to order 173, thus more than doubling the number of terms 
obtained previously. In the following I shall very briefly describe
how the series are calculated. The actual algorithm is a simple
specialization of the one used for the bulk problem, and details
can be found in \cite{Jensen99}. The results of the analysis of the 
series are presented in Section~\ref{sec:analysis}.

\section{Series expansions \label{sec:serexp}}

In the low-density phase ($p<p_c$) many quantities of interest 
can be derived from the pair-connectedness $C_{x,t}(p)$, which
is the probability that the site $x$ is occupied at time
$t$ given that the origin was occupied at $t=0$. The moments
of the pair-connectedness diverge as $p$ approaches the critical point
from below:

\begin{equation}
\mu_{n,m}(p) = \sum_{t=0}^{\infty}\sum_{x} x^nt^m C_{x,t}(p)
\propto (p_c-p)^{-(\gamma_1+n\nu_{\perp,1}+m\nu_{\parallel,1})},
 \;\;\;\; p \rightarrow p_c^-.
\label{eq:moments}
\end{equation}

\noindent
In particular the average cluster size $S(p)=\mu_{0,0}(p) 
\propto (p_c-p)^{-\gamma}$. Any directed path to a site whose parallel 
distance from the origin is $t$ contains at least $t$ bonds. From this it 
follows that if $C_{x,t}$ has been calculated for $t \leq t_{\rm max}$ then 
one can determine the moments to order $t_{\rm max}$. The pair-connectedness 
can be calculated via a transfer-matrix type algorithm by moving a boundary 
line through the lattice one row at a time with each row built up one site 
at a time. The sum over all contributing graphs is calculated as the 
boundary line is moved through the lattice. At any given stage the 
boundary line cuts through a number of, say,
$k$ sites where each site $j$ is in a state $\sigma_{j}=1$ if there is a 
bond entering the site from the row above, and $\sigma_{j}=0$ otherwise.
Configurations along the boundary line can thus be represented as binary 
numbers, and the contribution from each configuration is given by a 
truncated polynomial in $p$. Let 
$S_{a,b}=(\sigma_1,....,\sigma_{j-1},a,b,\sigma_{j+2},...,\sigma_k)$ 
be the configuration of sites along the boundary with $\sigma_{j}=a$ and
$\sigma_{j+1}=b$. As the boundary is moved at position $j$, the 
boundary polynomials, $BP$ are updated as follows \cite{Jensen99}:

\begin{eqnarray*}
BP(S_{1,1}) & = & p^2BP(S_{1,0})+(p-p^2)BP(S_{1,1}) \\
BP(S_{0,1}) & = & pBP(S_{1,0})-pBP(S_{1,1})+BP(S_{0,1}) \\
BP(S_{1,0}) & = & pBP(S_{1,0}) \\
BP(S_{0,0}) & = & BP(S_{0,0}) \\
\end{eqnarray*}

In a calculation to a given order $N$ we need to calculate the
contributions for all $t \leq N$. For a
given $t' < t$ the possible configurations along the boundary line are
limited by constraints arising from the facts that graphs have
to terminate at level $t$ and have no dangling parts. 
The ``no dangling parts'' restraint is equivalent to demanding that
sites with incoming bonds also have outgoing bonds. Therefore
a configuration for which the maximal separation between sites
with incoming bonds is $r$ will take at least another $r$ steps
before collapsing to a configuration with a single incoming bond.
Consequently if $t'+r >  t$, that configuration makes no contribution to 
$C(x,t)$ for any $x$, and so can be discarded. Furthermore if the minimal 
order to which a configuration contributes, $N_{\rm cont} > N$,
the configuration can be discarded since it will only contribute at an 
order exceeding that to which we want to carry out our calculation.
In order to calculate $N_{\rm cont}$, observe that a configuration
can be constructed and deconstructed in this manner. First take 
$t'-r$ steps, start branching for $r$ steps until the given configuration 
is produced, then start coalescing branches for another $r$ steps, and
then take the remaining $t-t'-r$ steps. It is easy to calculate the minimum 
order, $N_{\rm min}$, of the boundary polynomial as the configuration is 
built up, and from the arguments given above it follows that

\begin{equation}
N_{\rm cont}=2N_{\rm min}+t-2t'.
\end{equation}

\noindent
Further memory savings are obtained by observing that in calculating  
$C(x,t)$ we know that the graphs have at least $t$ bonds so we need only 
store $N-t$ coefficients, and when the boundary is moved from one row to 
the next we discard the lowest order term in each boundary polynomial.

\section{Analysis of the series \label{sec:analysis}}

The series for moments of the pair-connectedness were analyzed 
using differential approximants. A comprehensive review of
these and other techniques for series analysis may be found in
\cite{Guttmann89}.  This allows one to locate the critical point 
and estimate the associated critical exponents fairly accurately, 
even in cases  where there are additional non-physical singularities.  
Here it suffices to say that a $K$th-order
differential approximant to a function $f$, for which one has derived
a series expansion, is formed by matching the coefficients in the
polynomials $Q_i$ and $P$ of order $N_i$ and $L$, respective,
so that the solution to the inhomogeneous differential equation
\begin{equation}\label{eq:diffapp}
\sum_{i=0}^K Q_{i}(x)(x\frac{\mbox{d}}{\mbox{d}x})^i \tilde{f}(x) = P(x)
\end{equation}
agrees with the first series coefficients of $f$. The equations are
readily solved as long as the total number of unknown coefficients in
the polynomials is smaller than the order of the series $N$.
The possible singularities of the series appear as
the zeroes $x_i$ of the polynomial $Q_K$ and the associated critical
exponent $\lambda_i$ is estimated from the indicial equation

$$
\lambda_i=K-1-\frac{Q_{K-1}(x_i)}{x_iQ_K ' (x_i)}.
$$
The physical critical point is generally the first singularity on
the positive real axis.

\subsection{The critical point and exponents}

In this section I will give a detailed account of the results of the
analysis of the series $S(p)$, $\mu_{1,0}(p)$, $\mu_{2,0}(p)$,
$\mu_{0,1}(p)$, and $\mu_{0,2}(p)$. For fixed $K$ and a fixed order, $L$, 
of the inhomogeneous polynomial, estimates of the critical point and 
exponents were obtained by averaging over differential approximants
using at least 150 terms, with the further constraint that the 
order of the polynomials  $Q_{j}$ differ by at most 1. 
In Table~\ref{table:crpexp} I have listed the estimates obtained
from second- and third-order differential approximants for a few
values of $L$. The numbers in parenthesis are error-estimates
reflecting the spread among the various approximants. It should
be emphasized that these errors are at best indicative of the
`true' error since they do not reflect more systematic errors
such as a possible systematic drift in the estimates as the order of
the polynomials used to form the approximants is increased. In short
the quoted errors will tend to be too small and give a false sense
of how well-converged the estimates really are.

Apart from a few second-order cases with low $L$-values
the estimates are consistent with $p_c=0.644700175(35)$. This is in
excellent agreement with the more accurate estimate 
$p_c=0.644700185(5)$ obtained from the bulk series \cite{Jensen99},
and confirms without reasonable doubt the observation made in \cite{EGJT} 
that the introduction of the wall does not change the value of the
critical point. It is worth noting that the estimates from the
series $\mu_{0,2}(p)$ are exceptionally close to the bulk estimate.
Note also that there is a small but systematic change in the
$p_c$ estimates from the remaining series as one goes from second-
to third-order approximants, with the latter being closer to the
bulk estimate. From this table one can obtain the following estimates
for the critical exponents

\begin{eqnarray*}
\gamma_1  &= & 1.820544(15),\\ 
\gamma_1+\nu_{\parallel,1} & = & 3.554385(25), \\
\gamma_1+2\nu_{\parallel,1} & = & 5.288204(4) , \\
\gamma_1+\nu_{\perp,1}  &= &  2.91725(4), \\
\gamma_1+2\nu_{\perp,1}  &= & 4.01406(5). 
\end{eqnarray*}

\noindent
Note that since these estimates takes into account the differences 
between the second- and third-order approximants, and the variation 
with $L$, the errors are likely to be conservative. In particular they
are typically an order of magnitude larger than the  `errors' on
any specific estimate in Table~\ref{table:crpexp}.
From the exponent estimates we find that $\nu_{\parallel,1}=1.733820(12)$
and $\nu_{\perp,1}=1.0968(1)$. The corresponding estimates for
the bulk problem are $\nu_{\parallel}  = 1.733847(6)$ and 
$\nu_{\perp}  =  1.096854(4)$, which confirms that 
$\nu_{\parallel,1}=\nu_{\parallel}$ and  $\nu_{\perp,1}=\nu_{\perp}$
as postulated in \cite{EGJT}. It is evident from Table~\ref{table:crpexp}
that the series $\mu_{1,0}(p)$ and $\mu_{2,0}(p)$ appear to be the
worst converged.  I therefore also analyzed the series 
$\mu_{2,0}(p)/\mu_{1,0}(p) \propto (p_c-p)^{-\nu_{\perp}}$
and found this to yield the much more accurate estimates
$p_c=0.644700185(8)$ and $\nu_{\perp}=1.09685(1)$, which don't
require further comment.

\begin{figure}
\begin{center}
\includegraphics{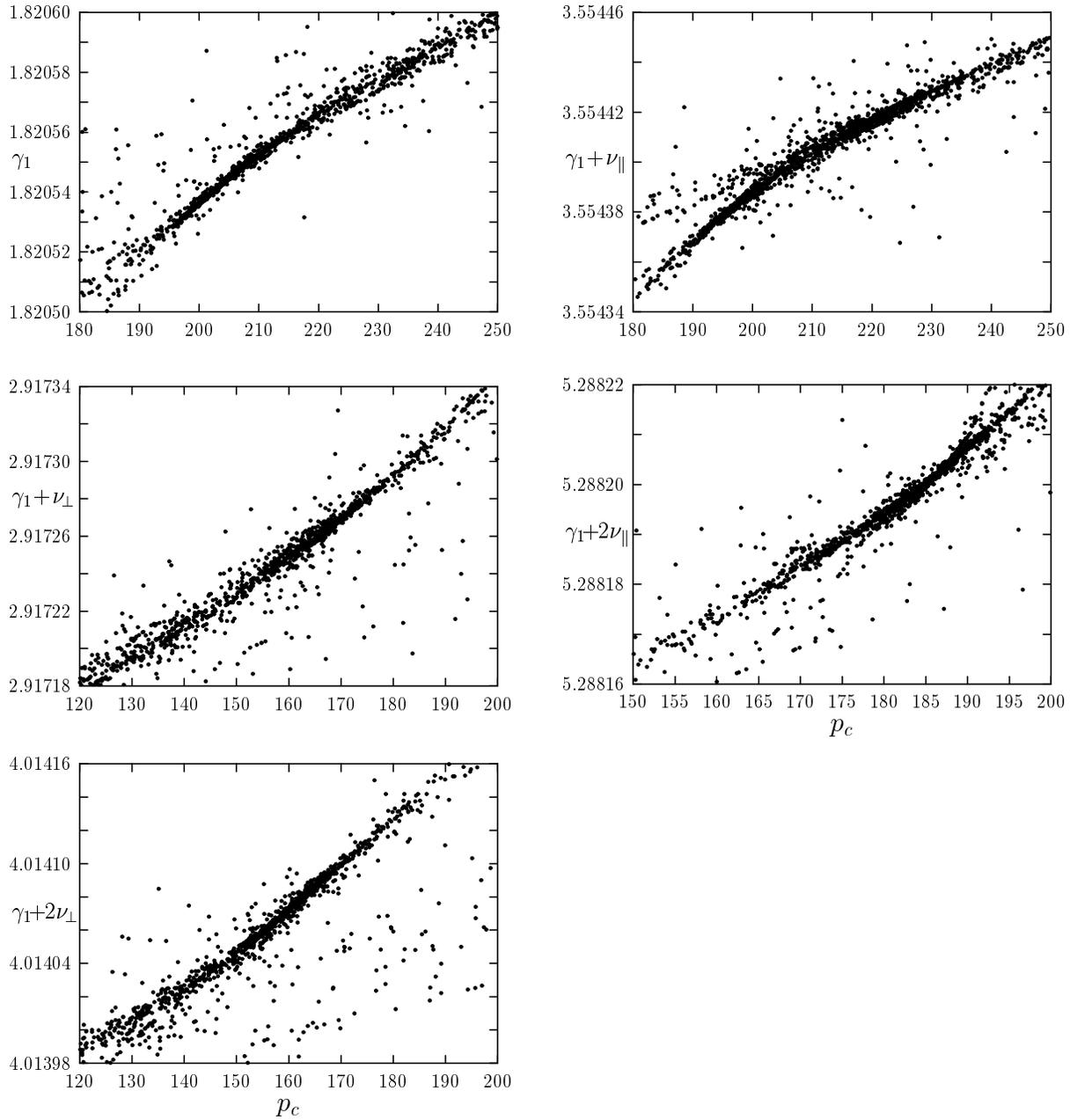}
\end{center}

\caption{\label{fig:crpexp} Estimates of the critical exponents
obtained from third-order differential approximants vs
estimates of the critical point. 
Numbers along the $x$-axis are all preceded by 0.644700.
Shown are (from left to right and top to bottom) estimates from
the series $S(p)$, $\mu_{0,1}(p)$, $\mu_{1,0}(p)$, $\mu_{0,2}(p)$,
and  $\mu_{2,0}(p)$.}

\end{figure}

Due to the high degree of internal consistency of the estimates from 
the bulk series one would tend to believe quite firmly in their accuracy 
and correctness and one can then use them to try and obtain more accurate 
estimate for the exponents for the problem with a wall. In 
Fig.~\ref{fig:crpexp} I have plotted the estimates for the critical 
exponents vs the estimates for $p_c$. By extrapolating the $p_c$ estimates 
until they lie in the interval given by the bulk estimate 
$p_c=0.644700185(5)$, one obtains the following `biased' estimates for the 
exponents for the wall problem

\begin{eqnarray}\label{eq:expest}
\gamma_1  &= & 1.82051(1),\nonumber\\ 
\gamma_1+\nu_{\parallel,1} & = & 3.55436(1),\nonumber \\
\gamma_1+2\nu_{\parallel,1} & = & 5.288202(6), \\
\gamma_1+\nu_{\perp,1}  &= & 2.917305(15), \nonumber \\
\gamma_1+2\nu_{\perp,1}  &= & 4.01414(1). \nonumber
\end{eqnarray}

\noindent
These estimates have a very high degree of consistency and I therefore
take $\gamma_1  =  1.82051(1)$ as the final estimate. Again the errors
are likely to be conservative since they reflect the variation of the
exponents over the entire interval of the bulk estimate of $p_c$.

In order to check the validity of the conjecture $\tau_1=1$ one
has to rely on scaling relations to express $\tau_1$ in
terms of $\gamma_1$ and the bulk exponents $\gamma$, $\nu_{\parallel}$
and $\nu_{\perp}$. First we note that \cite{EGJT}

\begin{equation}
\tau_1 = \nu_{\parallel}-\beta_1,
\end{equation}
\noindent
where $\beta_1$ is the exponent characterizing the decay of the percolation 
probability. Next we use a hyper-scaling relation derived for the case with 
the seed close to the wall \cite{Frojdh} 
\begin{equation}
\nu_{\parallel}+d\nu_{\perp}=\beta+\beta_1+\gamma_1,
\end{equation}
\noindent
where $d+1$ is the dimension of the lattice. This is a simple generalisation 
of the usual bulk hyper-scaling relation
\begin{equation}
\nu_{\parallel}+d\nu_{\perp}=2\beta+\gamma.
\end{equation}

\noindent
By combining all of these relations one finds

\begin{equation}
\tau_1 = \gamma_1-(\gamma+d\nu_{\perp}-\nu_{\parallel})/2.
\end{equation}

\noindent
In \cite{Jensen99} it was estimated that $\gamma=2.277730(5)$ and
by inserting the previously stated estimates for the remaining exponents
one gets the estimate $\tau_1=1.00014(2)$. So this would clearly
rule out the possibility that  $\tau_1=1$, though it is tantalizingly
close.  

\subsection{Non-physical singularities}

The series have a radius of convergence smaller than $p_c$ due to 
singularities in the complex $p$-plane closer to the origin than the 
physical critical point. Since all the coefficients in the expansion 
are real, complex singularities always come in conjugate pairs. 
The analysis indicates that the series have quite a large number of 
non-physical singularities, namely a singularity on the negative
real axis and three conjugate pairs in the complex plane.
The singularity on the negative real axis at $p_-=-0.51666(1)$
and two of the  conjugate pairs at $p_1=0.01005(10)\pm 0.47495(15)i$ 
and $p_2=-0.2255(10)\pm 0.4395(10)i$ are also present in the bulk
series, while the third conjugate pair at $p_3=0.225(5)\pm 0.420(5)i$
is unique to the wall problem. At the singularity $p_-$ the exponent 
estimates are $0.533(3)$,  $-0.463(4)$,  $-1.443(3)$, $0.23(5)$, 
and $-0.10(3)$, obtained from the series for $S$, $\mu_{0,1}$, $\mu_{0,2}$, 
$\mu_{1,0}$, and $\mu_{2,0}$, respectively. The corresponding exponent
estimates at the conjugate pair of singularities $p_1$ are
$4.0(1)$, $2,98(5)$, $2.00(5)$, $4.0(3)$, and $4.0(5)$. No meaningful
estimates could be obtained for the exponents at $p_2$ and $p_3$.

\section{Conclusion \label{sec:conc}}

In this paper I have reported on the derivation and analysis of
low-density series for moments of the pair-connectedness on the
directed square lattice with the origin close to an impenetrable wall.
The series for bond percolation was extended to order 173 as compared to 
order 67 obtained in previous work \cite{EGJT}. Analysis of the series led 
to very accurate estimates for the critical parameters and clearly showed 
that the critical point $p_c$ and the exponents $\nu_{\parallel}$ and 
$\nu_{\perp}$ have the same values as in the bulk. However, the exponent
for the average cluster size differs from the bulk case and has the
value $\gamma_1=1.82051(1)$.

Using scaling relations an estimate, $\tau_1=1.00014(2)$, was obtained
for the exponent characterizing the average cluster length. This 
rules out the possibility that  $\tau_1=1$. This conclusion, however, 
hinges crucially on having accurate estimates for four exponents and 
they would not require much altering to get an estimate consistent 
with $\tau_1=1$. On the other hand, so far no one has been able to give 
theoretical arguments to support $\tau_1=1$. Since other exponents for 
this problem appear not even to be given by (simple) rational fractions,
the weight of evidence would at present clearly not favor the conjecture
$\tau_1=1$. A more direct confirmation of this, say from an extended
series for the average cluster length, would clearly be desirable.

\section*{E-mail or WWW retrieval of series}

The series can be obtained via e-mail by sending a request to 
I.Jensen@ms.unimelb.edu.au or via the world wide web on the URL
http://www.ms.unimelb.edu.au/\~{ }iwan/ by following the instructions.

\section*{Acknowledgments}
 
The work was supported by a grant from the Australian Research Council.

\eject

\begin{table}
\caption{\label{table:crpexp}
Estimates of the location of the critical point and exponents obtained from 
second and third-order differential approximants.}
\begin{center}
\begin{tabular}{rllll}  \hline  \hline

& \multicolumn{2}{c}{Second-order DA} &
\multicolumn{2}{c}{Third-order DA} \\
\hline
\multicolumn{1}{c}{$L$} &
\multicolumn{1}{c}{$p_c$} & \multicolumn{1}{c}{$\gamma_1 $} &
\multicolumn{1}{c}{$p_c$} & \multicolumn{1}{c}{$\gamma_1 $} \\
\hline
0  &  0.644700235(11)& 1.820594(27)&0.6447002034(41)& 1.8205419(69) \\
10  &  0.644700232(10)& 1.820582(32)& 0.6447002036(24)& 1.8205425(41)\\ 
20  &  0.6447002093(67)& 1.8205517(98)& 0.6447002012(15)& 1.8205381(26)\\
30  & 0.6447002094(48)& 1.8205509(68)&0.6447002021(29)& 1.8205396(50) \\
40  &  0.6447002072(20)& 1.8205486(35)& 0.6447002037(27)& 1.8205426(49)\\
50  & 0.6447002071(16)& 1.8205491(29)& 0.6447002057(20)& 1.8205461(37)\\
\hline
\multicolumn{1}{c}{$L$} &
\multicolumn{1}{c}{$p_c$} &
\multicolumn{1}{c}{$\gamma_1+\nu_{\parallel,1}$} &
\multicolumn{1}{c}{$p_c$} &
\multicolumn{1}{c}{$\gamma_1+\nu_{\parallel,1}$} \\
\hline
0 & 0.6447002058(13)& 3.554408(22)& 0.6447001967(10)& 3.5543816(24)  \\
10  &0.6447002051(22)& 3.554399(13)& 0.64470019619(94)& 3.5543805(21)\\
20  &0.6447002002(33)& 3.5543903(76)& 0.6447001971(17)& 3.5543826(38)\\
30  &0.6447002014(55)& 3.554392(16)& 0.6447001971(19)& 3.5543827(35)\\
40  &0.6447002031(14)& 3.554398(11)& 0.64470019670(80)& 3.5543815(15)\\
50  & 0.6447002047(58)& 3.554403(20)&  0.6447001961(29)& 3.5543809(61)\\
\hline
\multicolumn{1}{c}{$L$} &
\multicolumn{1}{c}{$p_c$} &
\multicolumn{1}{c}{$\gamma_1+2\nu_{\parallel,1} $} &
\multicolumn{1}{c}{$p_c$} &
\multicolumn{1}{c}{$\gamma_1+2\nu_{\parallel,1} $} \\
\hline
0  &0.64470018744(16)& 5.28820382(35)&  0.64470018740(20)& 5.28820376(35)\\
10  &0.64470018742(10)& 5.28820374(23)& 0.64470018728(15)& 5.28820356(26) \\
20  & 0.64470018719(40)& 5.28820332(70)&0.64470018744(46)& 5.28820389(87)\\
30  & 0.64470018761(46)& 5.28820417(93)&0.64470018752(22)& 5.28820400(41)\\
40  & 0.64470018782(22)& 5.28820461(41) &0.6447001877(10)& 5.2882045(23)\\
50  & 0.64470018744(61)& 5.2882039(11)& 0.6447001867(20)& 5.2882032(30) \\
\hline
\multicolumn{1}{c}{$L$} &
\multicolumn{1}{c}{$p_c$} &
\multicolumn{1}{c}{$\gamma_1+\nu_{\perp,1} $} &
\multicolumn{1}{c}{$p_c$} &
\multicolumn{1}{c}{$\gamma_1+\nu_{\perp,1} $} \\
\hline
0  &0.6447001398(82)& 2.917206(13)& 0.644700161(13)& 2.917255(35)\\
10  & 0.6447001397(48)& 2.917212(28)&  0.6447001628(68)& 2.917257(21)\\
20  &  0.6447001498(67)& 2.917206(10)&0.6447001675(18)& 2.9172640(40)\\
30  & 0.6447001513(47)& 2.9172274(76)&0.644700176(11)& 2.917286(29)\\
40  & 0.6447001479(27)& 2.917221(19)&0.6447001701(32)& 2.9172700(76)\\
50  &0.6447001478(51)& 2.917229(13)& 0.6447001735(88)& 2.917277(21)\\
\hline
\multicolumn{1}{c}{$L$} &
\multicolumn{1}{c}{$p_c$} &
\multicolumn{1}{c}{$\gamma_1+2\nu_{\perp,1} $} &
\multicolumn{1}{c}{$p_c$} &
\multicolumn{1}{c}{$\gamma_1+2\nu_{\perp,1} $} \\
\hline
0  & 0.644700140(10)& 4.014024(20)&0.6447001577(35)& 4.014066(10) \\
10  & 0.6447001408(75)& 4.014025(16)& 0.6447001575(46)& 4.014060(44)\\
20  & 0.644700142(11)& 4.014028(21)&0.6447001613(23)& 4.0140763(75)\\
30  & 0.644700126(12)& 4.01404(11)& 0.6447001636(25)& 4.0140813(69)\\
40  & 0.644700140(11)& 4.014025(53)&0.6447001710(68)& 4.014101(20)\\
50  & 0.6447001355(65)& 4.013985(41)& 0.644700149(10)& 4.014049(24)\\
\hline \hline
\end{tabular}
\end{center}
\end{table}

\eject

\end{document}